\documentclass[12pt]{article}
\usepackage[dvips]{color}
\usepackage{epsfig}
\usepackage{amsmath}
\usepackage{graphicx}
\def\Box{\hbox{$\rlap{$\sqcup$}\sqcap$}}
\begin{document}
\title{\bf Cosmic acceleration and  crossing of $\omega=-1$  in non-minimal
modified Gauss-Bonnet gravity }
\author{J. Sadeghi $^{a}$\thanks{Email: pouriya@ipm.ir}\hspace{1mm}
,M. R. Setare $^{b}$ \thanks{Email: rezakord@ipm.ir}\hspace{1mm} and
 A. Banijamali $^{a}$\thanks{Email: abanijamali@umz.ac.ir}\hspace{1mm}  \\
$^a$ {\small {\em  Sciences Faculty, Department of Physics, Mazandaran University,}}\\
{\small {\em P .O .Box 47415-416, Babolsar, Iran}}\\
$^{b}${\small {\em Department of Science, Payame Noor University, Bijar, Iran }}\\
} \maketitle
\begin{abstract}
\noindent \hspace{0.35cm}  \\
Modified Gauss-Bonnet, i.e, $f(G)$ gravity is a possible explanation
of dark energy. Late time cosmology for the $f(G)$ gravity
non-minimally coupled with a free massless scalar field have been
investigated in Ref. [32]. In this paper we generalize the work of
Ref. [32] by including scalar potential in the matter Lagrangian
which is non-minimally coupled with modified Gauss-Bonnet gravity.
Also we obtain the conditions for having a much more amazing problem
than the acceleration of the universe, i.e: crossing of $\omega=-1$,
in $f(G)$ non-minimally coupled with tachyonic Lagrangian.


\end{abstract}
\section{Introduction}
Recent accelerated expansion of our universe is one the most
significant cosmological discoveries over the last decade [1-4].\\
This acceleration is explained in terms of the so called dark
energy. Many candidates for the nature of dark energy has been
proposed. The simplest suggestion for dark energy is cosmological
constant. But it suffers from two kind of problems [5] : fine tuning
and coincidence problem. A dynamical scalar field with quintessence
or phantom behavior is
another proposal for dark energy ( for reviews see [6] ).\\
An alternative approach for the gravitational origin of dark energy
is coming from modification of general relativity (GR) [7-12]. The
simplest way for modification of GR is to replace Ricci scalar, $R$
in Einstein- Hilbert action with a general function of the Ricci
scalar which is well known as $f(R)$ gravity [13,14]. For this kind
of modification, one assumes that the gravitational action may
contain some additional terms which starts to grow with
decreasing curvature and obtain a late time acceleration epoch.\\
Alternatively one can consider the Lagrangian density as a general
function of the Ricci scalar field and the Gauss-Bonnet invariant,
$f(R,G)$ [12,15].\\
Gauss-Bonnet term arises naturally as the leading order of $\alpha'$
expansion of heterotic superstring theory, where, $\alpha'$ is the
inverse string tension [16], may lead to some interesting results in
cosmology. The four dimensional Gauss-Bonnet term is a topological
invariant and thus has no dynamical effect to the field equations if
added into the action linearly. To get some contribution in four
dimensional space-time, one may couple the Gauss-Bonnet term to a
scalar field, as naturally appears in low energy effective action of
the string theory [17,18]. The consequences of GB coupling with
scalar field to the late time universe have been studied in [19].
Other aspects of GB gravity coupled with scalar field such as
avoidance of naked singularities in dilatonic brane world scenarios
and the problem of
fine tuning have been discussed in [20].\\
In the other side the recent analysis of the type Ia supernovas data
indicates that dark energy equation of state parameter $\omega$
(defined as ratio of pressure to energy density) crosses -1 at
$\emph{z}=0.2$ from above to below [21], where $\omega=-1$ is the
equation of state for the cosmological constant. The dark energy
with $\omega<-1$ is called phantom dark energy [22], and phantom
scalar field has negative kinetic term. However, the equation of
state of phantom scalar field is always less than -1 and can not
realize crossing of -1. Also quintessence dark energy with
$\omega>-1$ is not suitable to make the equation of state cross over
the $\omega=-1$. In general it has been proved [23,24], that in
single scalar field models  crossing of the cosmological constant
boundary is impossible.  \\
A possible solution to this problem is hybrid model , composed of
two scalar fields (quintessence and phantom), and proposed by Ref.
[25]. Also there exist other models to solve this problem such as,
non-minimal scalar tensor models [26], hessence [27],
string-inspired models [28] and models including higher order
curvature invariant terms [29]. The possibility of crossing the
cosmological constant boundary through Gauss-Bonnet
interaction has been addressed in [30,31].\\
In this paper we use the model proposed in  Ref.[32], for dark
energy and late time cosmology. However we include the scalar field
potential in the matter-like Lagrangian and show that cosmic
acceleration may occur in such model for special choices of scalar
field and potential. Also we consider non-minimal coupling of $f(G)$
with tachyon field Lagrangian and discuss  the conditions require
for crossing of the cosmological constant boundary in such theory.\\

\section{Late time acceleration in modified Gauss-Bonnet gravity }
 We start with the following action [32],
\begin{equation}
S=\int d^{4}x \sqrt{-g}\Big[\frac{1}{2\kappa^{2}}R+f(G)L_{d}\Big].
\end{equation}
Here $L_{d}$ is dark energy Lagrangian and $G$ is Gauss-Bonnet
invariant:
\begin{equation}
G=R^{2}-4R_{\mu\nu}R^{\mu\nu}+R_{\mu\nu\rho\sigma}
R^{\mu\nu\rho\sigma},
\end{equation}
By variation over metric $g_{\mu\nu}$ one gets:
\begin{eqnarray}\label{eq1}
0&=&\frac{1}{2\kappa^{2}}\big(R_{\mu\nu}-\frac{1}{2}g_{\mu\nu}R\big)+2f'(G)L_{d}R\,R_{\mu\nu}\nonumber\\
&-&4f'(G)L_{d}R_{\mu}\,^{\rho}R_{\nu\rho}+2f'(G)L_{d}R_{\mu\rho\sigma\lambda}R_{\nu}\,^{\rho\sigma\lambda}
+4f'(G)L_{d}R_{\mu\rho\sigma\nu}R^{\rho\sigma}+2\big(\big(g_{\mu\nu}\Box-\nabla_{\mu}\nabla_{\nu}\big)
f'(G)L_{d}\big)R\nonumber\\
&+&4\big(\nabla^{\rho}\nabla_{\mu}
f'(G)L_{d}\big)R_{\nu\rho}+4\big(\nabla^{\rho}\nabla_{\nu}
f'(G)L_{d}\big)R_{\mu\rho}+4\big(\Box
f'(G)L_{d}\big)R_{\mu\nu}-4\big(g_{\mu\nu}\nabla^{\lambda}
\nabla^{\rho}f'(G)L_{d}\big) R_{\lambda\rho}\nonumber\\
&+&4\big(\nabla^{\rho}
\nabla^{\lambda}f'(G)L_{d}\big)R_{\mu\rho\nu\lambda}+f(G)T_{\mu\nu}.
\end{eqnarray}
where,
\begin{eqnarray}
T_{\mu\nu}=\frac{2}{\sqrt{-g}}\frac{\delta}{\delta
g^{\mu\nu}}\Big(\int d^{4}x\sqrt{-g}L_{d}\Big)
\end{eqnarray}
For a flat Friedman-Robertson-Walker (FRW) space-time with the
metric,
\begin{eqnarray}
ds^{2}=-dt^{2}+a^{2}(t)(dr^{2}+r^{2}d\Omega^{2})
\end{eqnarray}
$(0,0)$ component  of above equation (\ref{eq1}) has the following
form ,
\begin{eqnarray}
0=-\frac{3}{\kappa^{2}}H^{2}+Gf'(G)L_{d}-24H^{3}\frac{d}{dt}(f'(G)L_{d})+f(G)\rho_{m}
\end{eqnarray}
where $f'(G)=\frac{df(G)}{dG}$.\\
If we consider free massless scalar field
\begin{eqnarray}
L_{d}=-\frac{1}{2}g^{\mu\nu}\partial_{\mu}\varphi\partial_{\nu}\varphi
\end{eqnarray}
 then the equation of motion for
scalar field is as follows,
\begin{eqnarray}
0=\frac{1}{\sqrt{-g}}\frac{\delta S}{\delta \varphi}=
\frac{1}{\sqrt{-g}}\partial_{\mu}(f(G)\sqrt{-g}g^{\mu\nu}\partial_{\nu}\varphi).
\end{eqnarray}
We assume  $\varphi$ depends only on time $t$, therefore
\begin{eqnarray}
\dot{\varphi}=\frac{c}{a^{3}f(G)}
\end{eqnarray}
where $c$ is a constant.
 Now  we choose the
following function for $f(G)$ ,
\begin{equation}
f(G)=f_{0}G^{n}
\end{equation}
with constants $f_{0}$ and $n$. Substituting (9) and (10) in (6),
yield to;
\begin{eqnarray}
0&=&-\frac{3}{\kappa^{2}}H^{2}+\frac{288
c^{2}}{a^{6}f_{0}(\dot{H}H^{2}+H^{4})^{n+2}}\nonumber\\
&\times&\left\{(1+7n)H^{8}(4n^{2}+12n+2)\dot{H}H^{6}+n(n+1)\ddot{H}H^{5}+
(n+1)(2n+1)\dot{H}^{2}H^{4}\right\},
\end{eqnarray}
where $G$ in FRW space-time is $G=24(\dot{H}H^{2}+H^{4})$. \\
Now we are going to study power like solution of (11);
\begin{eqnarray}
a=a_{0}t^{x},\,\,\,\,\,\,\,\, H=\frac{x}{t},\,\,\,\,\,\,\,\,
\dot{H}=-\frac{x}{t^{2}},\,\,\,\,\,\,\,\, \ddot{H}=2\frac{x}{t^{3}}.
\end{eqnarray}
by inserting above relations into (11) one finds,
$x=\frac{2n+1}{3}$. Then universe accelerates, that is, $\ddot{a}>0$
if $n>1$ or $n<-\frac{1}{2}$.\\
From another side one can define the effective equation of state
parameter as
\begin{eqnarray}
\omega_{eff}=\frac{p}{\rho}=-1-\frac{2\dot{H}}{3H^{2}},
\end{eqnarray}
which in our case is $\omega=\frac{1-2n}{1+2n}$. So if
$n<-\frac{1}{2}$, we obtain an effective phantom while if $n>1$ the
effective quintessence occurs.\\
Let us take the case that the second term in (1) is dominant.
Assuming $H=\frac{x}{t}$, we find $x=\frac{(2n^{2}+6n+1)\pm
n\sqrt{16n^{2}-16n-6}}{(1+7n)}$, which is real if
$n>\frac{1}{2}(1+\frac{\sqrt{10}}{2})$, or
$n<\frac{1}{2}(1-\frac{\sqrt{10}}{2})$.\\
Now we generalize $L_{d}$ to include potential,
\begin{eqnarray}
L_{d}=-\frac{1}{2}g^{\mu\nu}\partial_{\mu}\varphi\partial_{\nu}\varphi-V(\varphi).
\end{eqnarray}
We try to find solution for the following forms of potentials and
scalar fields;
\begin{eqnarray}
V(\varphi)=V_{0}\varphi^{m},\,\,\,\,\,\,\,\,\varphi=\varphi_{0}t^{n},\,\,\,\,\,\,\,\,(a=a_{0}t^{x}),
\end{eqnarray}
where $V_{0}$ and $\varphi_{0}$ are constants.\\
After inserting (15) in (6) and assuming $f(G)$ as (10), we find
that, $m=2-\frac{2}{n}$.\\
Let us take again the case that the second term in (1) is dominant.
First, from equation of motion for $\varphi$,
\begin{eqnarray}
f(G)(\ddot{\varphi}+3H\dot{\varphi})+\dot{G}f'(G)=-f(G)V'(\varphi)
\end{eqnarray}
where $V'(\varphi)=\frac{dV(\varphi)}{d\varphi}$, one can obtain
\begin{eqnarray}
V(\varphi)=\frac{n\varphi}{2(1-n)}(\ddot{\varphi}+3H\dot{\varphi}+n\frac{\dot{G}}{G}).
\end{eqnarray}
Then for the solution (15) one gets $x=\frac{2n-3}{6(n-1)}$, which
shows cosmic acceleration if $\frac{3}{4}<n<1$ with $x>0$. The
corresponding  effective equation of state parameter is
$\omega_{eff}=\frac{2n-1}{2n-3}$, and we obtain an effective phantom
if $\frac{1}{2}<n<1$.
\section{The $\omega=-1$ crossing with tachyon field  }
Now we are going to explore crossing of  the  cosmological constant
barrier in the model represent by the action (1) but here scalar
field is tachyonic which has the following Lagrangian;
\begin{eqnarray}
L_{d}=-V(T)\sqrt{1+g^{\mu\nu}\partial_{\mu}T\partial_{\nu}T}
\end{eqnarray}
where $T$ is tachyon field and $V(T)$ is tachyonic potential.\\
In a flat FRW background  with the metric (5), and a homogenous
scalar field $T$, the equation of motion can be written as,
\begin{eqnarray}
\ddot{T}+3H\dot{T}=-\frac{V_{T}(T)}{V(T)}+
\frac{\dot{T}^{2}\ddot{T}}{1-\dot{T}^{2}}-24\frac{f'(G)}{f(G)}(\ddot{H}H^{2}+2\dot{H}^{2}H+4\dot{H}H^{3})\dot{T},
\end{eqnarray}
where $V_{T}(T)=\frac{dV(T)}{dT}$.\\
One can find the energy density and pressure generated by the scalar
field and modified Gauss-Bonnet gravity as follows;

\begin{eqnarray}
\rho=-Gf'V\sqrt{1-\dot{T}^{2}}+24H^{3}\frac{d}{dt}
(f'V\sqrt{1-\dot{T}^{2}})+\frac{fV}{\sqrt{1-\dot{T}^{2}}},
\end{eqnarray}

\begin{eqnarray}
p=Gf'V\sqrt{1-\dot{T}^{2}}-\frac{2G}{3H}\frac{d}{dt}
(f'V\sqrt{1-\dot{T}^{2}})-8H^{2}\frac{d^{2}}{dt^{2}}(f'V\sqrt{1-\dot{T}^{2}})-f
V\sqrt{1-\dot{T}^{2}}
\end{eqnarray}
\\
We now study the cosmological evolution of EoS for the present
model.
 From equations (20) and (21) we have the following expression,
\begin{eqnarray}
\rho+p&=&-16H\dot{H}\frac{d}{dt}
(f'V\sqrt{1-\dot{T}^{2}})+8H^{3}\frac{d}{dt}
(f'V\sqrt{1-\dot{T}^{2}})\nonumber\\
 &-&8H^{2}\frac{d^{2}}{dt^{2}}
(f'V\sqrt{1-\dot{T}^{2}})+\frac{fV\dot{T}^{2}}{\sqrt{1-\dot{T}^{2}}}
\end{eqnarray}
Since $\rho+p=(1+\omega)\rho$, one needs $\rho+p=0$ when
$\omega\longrightarrow-1$. To check the possibility of the crossing
of the phantom divide line $\omega=-1$, we have to explore for
conditions that $\frac{d}{dt}(\rho+p)\neq 0$ when $\omega$ crosses
over -1.  If we assume $\dot{T}=0$   when $\omega$ crosses -1 then
from Eq. (22) at the crossing point, one obtains;
\begin{eqnarray}
\rho+p=8H\dot{f'}V(H^{2}-2\dot{H})-8H^{2}\left(\ddot{f'}V+f'\ddot{T}(V_{T}-V\ddot{T})\right),
\end{eqnarray}
and
\begin{eqnarray}
\frac{d}{dt}(\rho+p)&=&-8\dot{f'}V\left(2\dot{H}^{2}+2H\ddot{H}-3\dot{H}H^{2}\right)\nonumber\\
&+&8H(H^{2}-4\dot{H})\left(\ddot{f'}V+f'\ddot{T}(V_{T}-V\ddot{T})\right)\nonumber\\
&-&8H^{2}\left(\dddot{f'}V+3\dot{f'}\ddot{T}(V_{T}-V\ddot{T})+f'\dddot{T}(V_{T}-3V\ddot{T})\right).
\end{eqnarray}
Now considering crossing point as a de Sitter point lead to
vanishing of all time derivatives of $f'(G)$ and from equation [23]
one gets $V_{T}=V\ddot{T}$. Then in order to equation [24] become
non-zero, we must have $\dddot{T}\neq$. So one can express the
conditions require for $\omega$ crossing -1 at a de Sitter point as:
$\dot{T}=0$, $\ddot{T}\neq0$ and $\frac{d}{dt}\Box T\neq0$.
Therefore crossing over $\omega=-1$ must happen before the tachyon
potential reaching it's minimum asymptotically. This is in agreement
with the result of Ref. [33] where the authors have obtained the
same result by adding a higher order derivative operator $\Box T$ in
the Lagrangian of tachyon field. But we obtained these result by
considering the non-minimal coupling of tachyon Lagrangian with
$f(G)$ gravity.
\section{Conclusion}
The recent cosmological observations indicate that the accelerated
expansion of the universe can be drive by dark energy component with
equation of state across -1. In the present work we considered a
non-minimally coupled modified Gauss-Bonnet gravity. The non-minimal
coupling parameter was assumed to be Lagrangian of a scalar field
which is included kinetic term and potential as well. For the
solutions (15) it was shown that the universe accelerates if
$\frac{3}{4}<n<1$ and we obtain an effective phantom epoch if
$\frac{1}{2}<n<1$.\\
Also we obtained the condition for crossing of $\omega=-1$ in $f(G)$
non-minimally coupled with tachyonic Lagrangian. We showed  that the
crossing over $\omega=-1$ must happen before the tachyon potential
reaching it's minimum asymptotically.

\end{document}